\documentstyle[epsfig,psfig,11pt]{article}

\def\beq{\begin{equation}}
\def\eeq{\end{equation}}
\def\bei{\begin{itemize}}
\def\eei{\end{itemize}}
\def\bev{\begin{verbatim}}
\def\eev{\end{verbatim}}

\def\o2{0$_{2}$}

\def\microk{$\mu{\mbox{K}}$}

\def\ruo2{RuO$_{2}$}

\def\mathrelfun#1#2{\lower3.6pt\vbox{\baselineskip0pt\lineskip.9pt
  \ialign{$\mathsurround=0pt#1\hfil##\hfil$\crcr#2\crcr\sim\crcr}}}

\def\fun#1#2{\lower3.6pt\vbox{\baselineskip0pt\lineskip.9pt
  \ialign{$\mathsurround=0pt#1\hfil##\hfil$\crcr#2\crcr\sim\crcr}}}
\newcommand{\wisk}[1]{{\ifmmode{#1}\else{$#1$}\fi}}

\newcommand{\bfS}{\wisk{\mbox{\boldmath $S$}}}

\begin{document}

\bibliographystyle{plain}

\title{Polarization of the Atmosphere as a Foreground for 
Cosmic Microwave Background Polarization Experiments}

\author{Shaul Hanany$^{1}$, Philip Rosenkranz$^{2}$ \\
$^{1}$ The University of Minnesota, Twin Cities, Minneapolis, Minnesota \\
$^{2}$ Massachusetts Institute of Technology, Cambridge, Massachusetts }

\maketitle

\section*{Abstract}

We quantify the level of polarization of the atmosphere due to Zeeman
splitting of oxygen in the Earth's magnetic field and compare it to
the level of polarization expected from the polarization of the cosmic
microwave background radiation. The analysis focuses on the effect at
mid-latitudes and at large angular scales.  We find that from
stratospheric balloon borne platforms and for observations near
100~GHz the atmospheric linear and circular polarized intensities is
about $10^{-12}$ and $100 \times 10^{-9}$ K, respectively, making the
atmosphere a negligible source of foreground. From the ground the
linear and circular polarized intensities are about $10^{-9}$ and $100
\times 10^{-6}$ K, making the atmosphere a potential source of
foreground for the CMB E (B) mode signal if there is even a 1\%
(0.01\%) conversion of circular to linear polarization in the
instrument.

\section{Introduction}

Ground and balloon borne experiments are attempting to
detect and characterize the cosmic microwave background (CMB)
polarization signal.  These efforts will continue into the foreseeable
future and their success depends on a thorough understanding of the
sources of foreground emission.  The atmosphere is a source of
foreground emission because the magnetic field of Earth causes a
Zeeman-splitting of the energy levels of oxygen and emission due to
transitions between these levels occurs at the frequency band where
the CMB is most intense.  The Zeeman transitions are 100\% polarized
and it is therefore necessary to quantify to what extent polarized
emission and absorption in the atmosphere affects 
CMB polarization experiments.

Tinkham and Strandberg \cite{tinkham,tinkham2} and Townes and Schawlow
\cite{townes} have derived the energy levels of the ground state of the 
oxygen molecule with and without the presence of a magnetic
field, Lenoir \cite{lenoir}, and Rosenkranz and Staelin 
\cite{rosenkranz} have discussed polarized emission of oxygen 
in the Earth's magnetic field. 
Keating et al.\ \cite{keating} found that at a frequency of 30 GHz the
fraction of linear polarization due to this emission should be less
than $10\times10^{9}$ K.  In this paper we give concrete
predictions (rather than an upper limit) for the level of polarization
due to the atmosphere over a broad range of frequencies and 
compare this level to the signals expected from the E and B modes
of the CMB polarization.

\section{Atmospheric Polarization}

\begin{figure}[t]
\centerline{\psfig{file=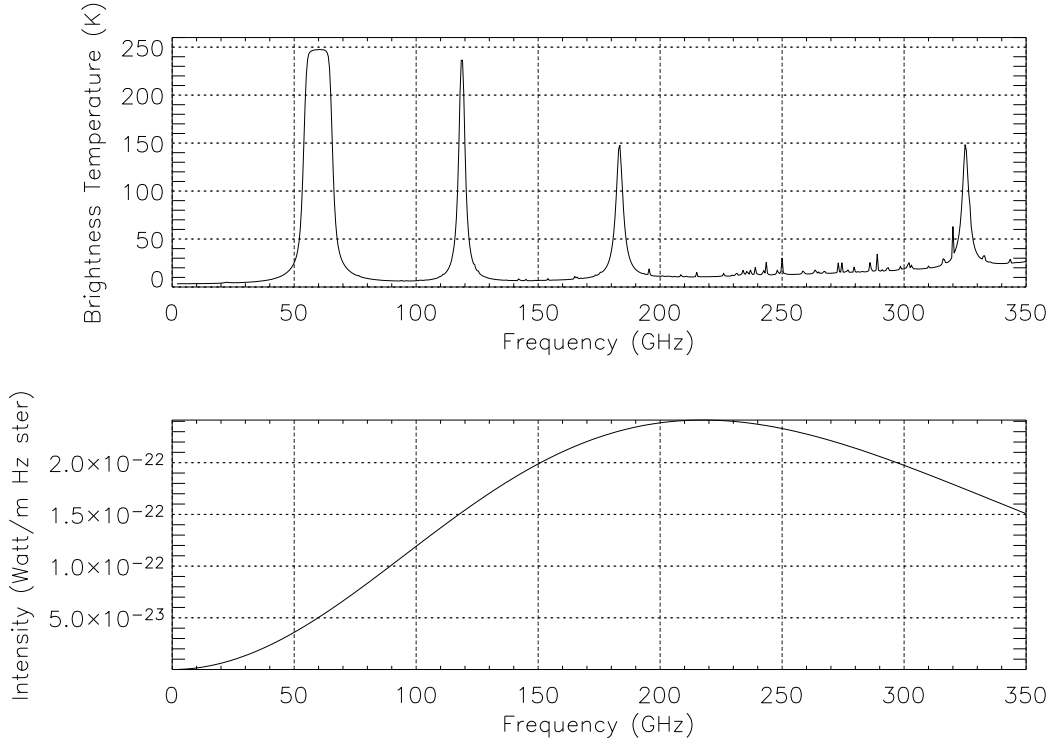,width=4.0in,angle=+90}}
\caption{The zenith brightness temperature of the US standard 
atmosphere \cite{standard_atmosphere} at an altitude of 5~km 
and relative humidity of 10\% (top panel)
and the spectrum of the fluctuations in the cosmic microwave background
radiation for an RMS fluctuation of 50 \microk\
(bottom panel), both as a function of frequency. Atmospheric 
windows near 30, 90 and 150~GHz make observations of the CMB attractive
at these frequency bands. }
\label{fig:spectrum.and.windows}
\end{figure}

Figure \ref{fig:spectrum.and.windows} shows the spectrum of the 
fluctuations of the CMB near its peak and the zenith effective temperature of the 
US Standard Atmosphere \cite{standard_atmosphere} at a 5 km high site. 
There are strong oxygen lines at $\sim 60$ and 118.8~GHz, a water line 
near 180 and 325~GHz, and a number of lower intensity ozone lines.  
Of these three molecules, oxygen is the only molecule that exhibits 
Zeeman splitting and has polarized emission. 

\begin{figure}[t]
\centerline{\psfig{file=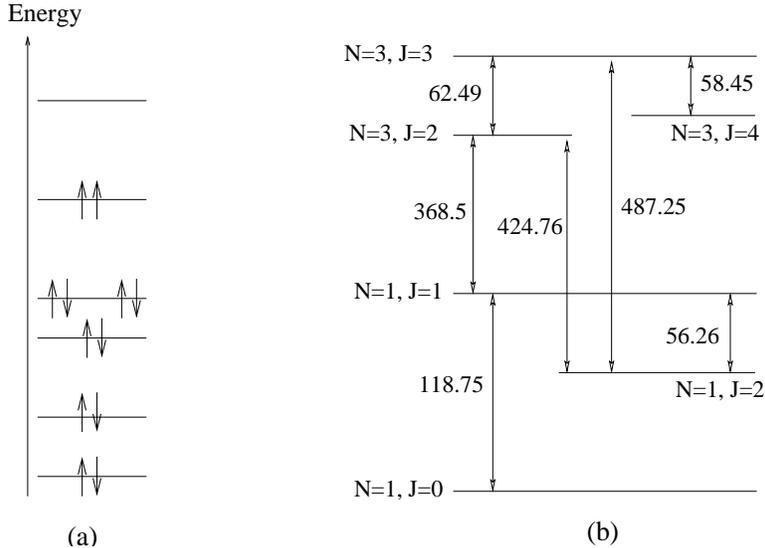,width=4.0in,angle=0}}
\caption{(a) The energy and spin configuration of the 12 valence electrons
of the oxygen molecule in its ground state. The two highest energy electrons,
which are in the $ 2\pi_{g}$ orbital, 
have their spins parallel giving a net spin $S=1$ to the molecule. The lower 
energy orbitals, which are in order the $1 \pi_{g}, \, 3\sigma_{g}, \, 2\sigma_{u}, \,$ and 
$ 1\sigma_{u}$, and the 4 inner shell electrons (not shown) give a net spin of zero.  
(b) The first two rotational energy levels, N=1 and 3, of the oxygen molecule 
in the electronic and vibrational ground state. The N and J pairs give the rotational 
and total angular momentum quantum numbers. The numbers near the arrows give the 
frequency of the transition in GHz. The Zeeman splitting is not shown.  }
\label{fig:energy_states}
\end{figure}

At its electronic ground state the oxygen molecule has 12 valence
electrons, as shown in Fig.~\ref{fig:energy_states}.  The two
electrons at the highest energy level pair up with their spins
parallel, while the spins of all other electrons combine to give spin
zero.  Therefore the total spin quantum number of the molecule is $S =
1$.  At the mm-wave frequency band the transitions of the oxygen
molecule are between rotational states of the electronic and
vibrational ground state.  The rotational angular momentum, with a
quantum number $N$, couples with the spin to give a total angular
momentum state $J$.  The resulting energy levels and rules for the
allowed transitions between them are summarized by Lenoir
\cite{lenoir}.  Figure~\ref{fig:energy_states} shows the transitions
between the first two rotational levels, $N=1$ and 3. The 118.8 GHz
line is the transition between a $J=0$ and a $J=1$ states, both of them
with $N=1$. There are a number of transitions with frequencies around 60
GHz, which give the strong line between 55 and 65 GHz
shown in Fig.~\ref{fig:spectrum.and.windows}.

In the presence of the Earth's magnetic field the $2J+1$ degeneracy of
each of the $J$ states is broken giving rise to a Zeeman splitting of
the energy levels and therefore to many new transitions.  Expressions
for the magnitude of the energy split as a function of the quantum
numbers and the magnetic field strength and for the selection rules
for transitions between the states are given by Lenoir \cite{lenoir}. (Notice an
inconsistency in the notation of the energy splitting between Lenoir
and Townes and Schawlow \cite{townes}; we follow the notation of
Townes and Schawlow, which is the more prevalent in the literature.)
For the typical strength of the Earth magnetic field the frequency
shift of the transitions due to the Zeeman splitting is on the order
of 1 MHz.

\begin{figure}[t]
\centerline{\psfig{file=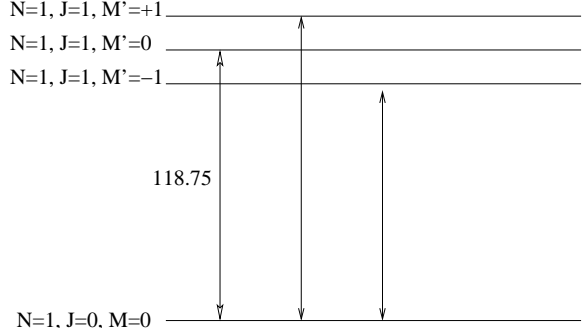,width=3.0in,angle=0}}
\caption{Zeeman splitting of the $(N,J)=(1,1)$ energy level to three levels
in the presence of a magnetic field. The $\Delta M = 0$ transition
is called $\pi$ and the $\Delta M = \pm 1$ transitions are called
$\sigma_{\pm}$, where  $ \Delta M \equiv M-M'$. }
\label{fig:zeeman}
\end{figure}

For simplicity we will focus our attention on the 118.8 GHz line,
although our analysis is valid for all of the oxygen lines.  In the
presence of a magnetic field the $(N,J)=(1,1)$ state is split to three states an
$M=-1$, 0, and +1 states ordered in increasing energy, as 
shown in Fig.~\ref{fig:zeeman}.  The transition with $\Delta M = 0$
is called the $\pi$ transition and the transitions with 
$\Delta M = \pm 1$, where $\Delta M = M_{low\,\,energy} - M_{high\,\, energy}$,
are called $\sigma_{\pm}$. These transitions are 100\% polarized and
their intensity, and the orientation of the polarization is
determined by the angle between the line of sight and the
magnetic field direction. 
When the line of sight is perpendicular to the magnetic field the 
two $\sigma$ lines are linearly polarized both in the same direction and  
parallel to the direction of the magnetic field, and the $\pi$ line
is linearly polarized perpendicular to the field. When the line of sight
is parallel to the magnetic field the $\pi$ line completely disappears
and the two $\sigma$ lines are circularly polarized in opposing orientations.  
A more detailed discussion of intermediate cases is given by Lenoir \cite{lenoir}. 
Radiation propagating through the atmosphere at frequencies near 120
GHz (or near any other oxygen line) will become polarized because of
differential polarized absorption and emission by the $\sigma$ and
$\pi$ lines.

\subsection{Order of Magnitude Estimate}

It is not difficult to give an order of magnitude estimate of the
polarized intensity of the 118~GHz oxygen line in emission due to the
Zeeman splitting. Since we are mainly interested in frequencies at
which the atmosphere is not opaque, we can neglect Doppler broadening.
We further assume that the line profile is collisional with no mixing,
that is Lorenzian. For the 118~GHz line this assumption is valid
because the relative strength of the three Zeeman components changes
both with polarization and viewing geometry.  Any mixing between two
transitions is reciprocal in the sense that it acts oppositely on
them, and isolation of one transition would imply a non-physical line
shape, such as negative absorption.  One Zeeman component can not
couple to another component that might disappear to a different
observer. It can only couple to another component of the same $\Delta
M$.  (See also the discussion following Eq.~3 of Rosenkranz and
Staelin
\cite{rosenkranz}.) Hence the emission from the three components is
incoherent and we simply sum the Stokes vectors that are associated
with each of them
\beq
\bfS (\nu, \theta) = \mu_{\pi} L_{\pi}(\nu)\bfS_{\pi}(\theta) + 
  \mu_{\sigma_{+}} L_{\sigma_{+}}(\nu)\bfS_{\sigma_{+}}(\theta) +
  \mu_{\sigma_{-}} L_{\sigma_{-}}(\nu)\bfS_{\sigma_{-}}(\theta),  
\eeq
where $\bfS$ is the appropriate Stokes vector, $\nu$ is the frequency of interest,
$\theta$ is the angle between the 
line of sight and the direction of the magnetic field, and 
$\mu$ is the relative strength of the transition, which is proportional 
to the square of the matrix element between the energy states. The Lorentzian functions 
are given by 
\beq
L(\nu) = I_{0} { \gamma \over (\nu - \nu_{c})^{2} + \gamma^{2} } 
\eeq
with 
\beq
\nu_{c} = \left\{ \begin{array}{ll}
\nu_{0} &  \mbox{for } \pi \\
\nu_{0}-\delta & \mbox{for } \sigma_{+} \\
\nu_{0}+\delta & \mbox{for } \sigma_{-} 
\end{array} \right. ,
\eeq
where $\nu_{0}$ is the frequency of the $\Delta M=0$ transition, $\delta$ is 
the magnitude of the shift in frequency, and $I_{0}$ is a normalizing constant 
proportional to the temperature of the atmosphere. 
The Stokes vectors of the lines are 
\beq
\bfS_{\sigma_{\pm}} =  
\left( \begin{array}{c} 
1+\cos^{2}\theta \\ 1-\cos^{2}\theta \\ 0 \\ \pm 2\cos\theta \end{array} 
\right)
\eeq
and 
\beq
\bfS_{\pi} =  \sin^{2} \theta
\left( \begin{array}{c} 
1 \\ -1 \\ 0 \\ 0 \end{array} 
\right),
\eeq
and $(\mu_{\pi},\mu_{\sigma_{+}},\mu_{\sigma_{-}}) = (2,1,1)$. 

We estimate the magnitude of the effect for two limiting cases
of observation parallel to and perpendicular to the direction of 
the magnetic field. 
In the limiting case of an observation 
parallel to the field, for which $\theta = 0$, we have 
\beq
\bfS_{\sigma_{\pm}} =  
\left( \begin{array}{c} 
2 \\ 0 \\ 0 \\ \pm 2 \end{array} 
\right)
\eeq
and the null vector for $ S_{\pi}$, which gives two circularly
polarized lines from the $\sigma$ components and no $\pi$ component.
For the case of observation perpendicular to the field, for
which $\theta = \pi/2$, we have
\beq
\bfS_{\sigma_{\pm}} =  
\left( \begin{array}{c} 
1 \\ 1 \\ 0 \\ 0 \end{array} 
\right)
\eeq
and 
\beq
\bfS_{\pi} = \left( \begin{array}{c} 
1 \\ -1 \\ 0 \\ 0 \end{array} 
\right)
\eeq
which gives orthogonal linear polarizations for the  
$\sigma$ and $\pi$ components. 

In the limit where $\gamma, \delta \ll (\nu-\nu_{0})$, which is 
always the case for altitudes of less than 50 Km above Earth, the 
polarized intensity for $\theta = \pi/2$ is purely a $Q$ Stokes state 
and is given by 
\beq
I_{p} (\nu,\theta=\pi/2) = Q(\nu) \simeq {1 \over 2} I_{t} 
\left( { \delta \over \nu - \nu_{0} } \right)^{2},
\eeq
where $I_{t}$ is the total intensity (the I Stokes parameter). 
The polarized intensity is quadratic in the ratio of $\delta$ to the
distance in frequency from line center $\nu - \nu_{0}$. for $\nu =
140$~GHz we have $\nu - \nu_{0} = 20$~GHz and with $\delta = 1$~MHz,
corresponding to a field of $\sim 1$~Gauss, and 
an atmosphere brightness temperature of
10 Kelvin we find $I_{p} \simeq 10\times 10^{-9}$ Kelvin, similar to
the estimate of Keating et al.\ \cite{keating}.

For the geometry where $\theta = 0$ the polarized intensity is purely a $V$ 
Stokes state, i.e. it is fully circular, but with a much larger amplitude 
\beq
I_{p} (\nu,\theta=0) = V(\nu) \simeq 4 I_{t} 
\left( { \delta \over \nu - \nu_{0} } \right).
\eeq
The polarized intensity is now linear in the ratio of $\delta$ to the 
distance in frequency from line center. for $\nu = 140$ GHz
we have $I_{p} \simeq 10^{-3}$ Kelvin. Even a 0.1\% conversion 
of circular polarization to linear in the instrument will give rise
to signals that are significant compared to the cosmological signals.  

\subsection{A More Exact Calculation}

The brightness temperature matrix of the atmosphere $\rm{T_{b}}$
at a frequency $\nu$ is given by \cite{lenoir}
\beq
\rm{T_{b}}(z) = \exp\left\{-\rm{G}(z)\right\} \rm{T_{b}}(0) \exp\left\{-\rm{G}^{*}(z)\right\}
        +\rm{T_{b}}(0) \left[ \rm{I} - \exp \left\{ -2\rm{A}(z) \right\} \right],
\eeq
where $\rm{G}(z)$ is the propagation matrix at a height $z$ above
Earth, $\rm{A}$ is the attenuation matrix and $\rm{I}$ is the identity
matrix.  We use the computer code of Rosenkranz and Staelin
\cite{rosenkranz} to calculate the polarized radiative transfer of the
cosmic microwave background radiation through two standard atmospheres
\cite{standard_atmosphere,humidity}: The 1976 US Standard Atmosphere,
which is characteristic for latitudes of about 40N with total water
vapor of 14 precip. mm, and a winter atmosphere at a latitude of 60N
with total water vapor of 4 precip. mm.  Magnetic field information
comes from the IGRF 1995 magnetic field model \cite{magnetic_field}.
Table ~\ref{tab:std76} gives the effective temperature of each of the
stokes parameters (in thermodynamic temperature) at three different
frequencies and two altitudes of observations. The observation sites
are at a longitude of 140~W and at either 40N or 60N, as appropriate,
and observations are pointed toward North at an elevation of 45
degrees. The calculations include contributions due to the 
60~GHz oxygen band, for which we calculate the line mixing using 
the parameters that were measured by Liebe et al. \cite{liebe2}. 
These measured values of line mixing (and line broadening) correspond
to the combination of Zeeman components within each line, and hence
are appropriate to the frequencies considered here, which lie
far outside the region of Zeeman structure. However, all 
of the frequencies considered here lie within 90 GHz of the line centers, 
so duration-of-collision effects are considered to be negligible. 


\section{Discussion}

\begin{figure}[t]
\centerline{\psfig{file=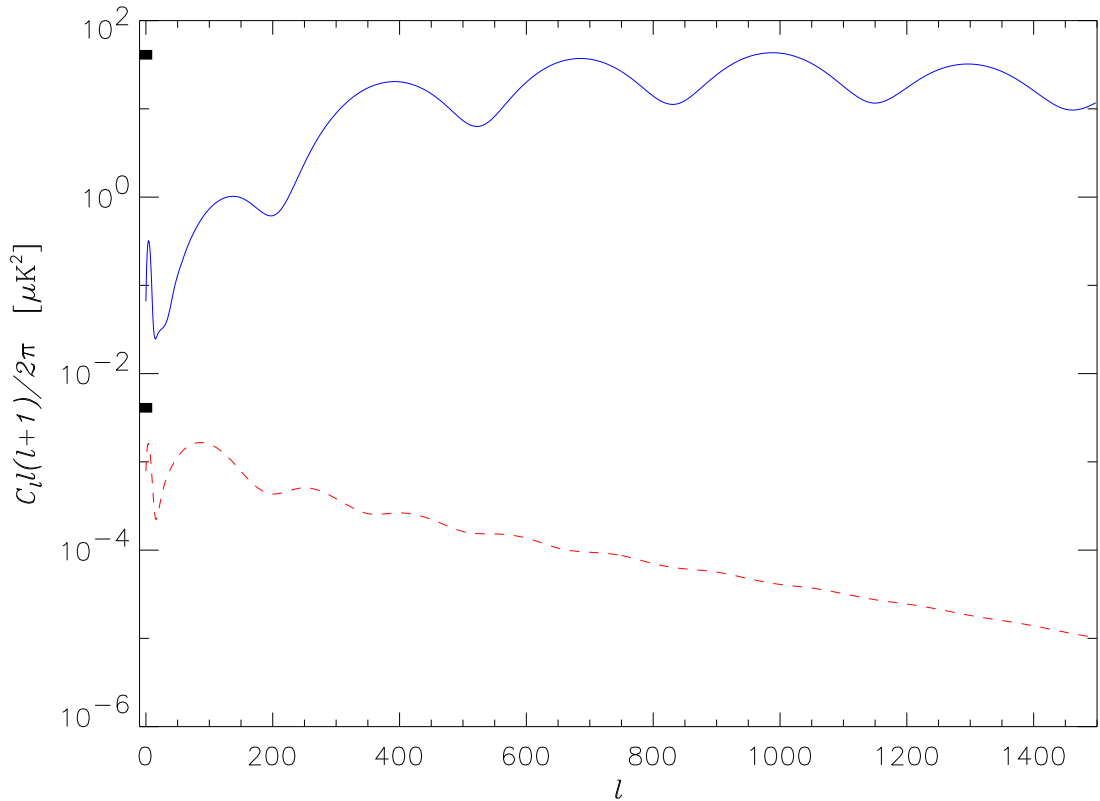,width=4.5in,angle=90}}
\caption{The angular power spectrum of the CMB E (solid) and B (dashed) 
mode polarization signals in a $\Lambda$CDM cosmology with 
tensor to scalar ratio $r = 0.01$. 
The top (bottom) short bars near $\ell=0$ show the 
level of atmospheric polarization at large angular scales that 
is expected for a 1\% (0.01\%) 
conversion of circular to linear polarization in an instrument
making observations from the ground at 90~GHz and at 
a latitude of 60N.  }
\label{fig:atm_and_cmb}
\end{figure}

The results of the calculations using the radiative transfer code and
the order of magnitude estimate generally agree and show that the
magnitude of the circular polarization is significantly larger than
that of the linear polarization component. Although the CMB itself is
not expected to be circularly polarized, a 1\% conversion of circular
to linear polarization in the instrument at a frequency of 90 GHz
would give rise to a $ 3 - 6 \times 10^{-6}$ K signal, depending on
the location of observation.  Such a signal is of the same order of
magnitude as the strongest 'E mode' signal and much larger than the
strongest 'B mode' signal expected from the CMB, as shown by the bars
in Figure~\ref{fig:atm_and_cmb}. However, the Figure also illustrates
that the spatial scale where the CMB peaks and the spatial scale that
is relevant for the calculation in this paper are quite distinct.
(Spatial scales are quantified in terms of $\ell \simeq 180/\theta$,
where $ \theta$ is the angle between two lines of sight in degrees.)
Under the assumption that the atmosphere is well mixed, an assumption
that underlies all atmospheric modeling, spatial variations of the
atmospheric polarization arise because of variations in the magnetic
field along different lines of sight.  However, the angular resolution
of the IGRF model provides description of spatial variations at $\ell
\simeq 1$ for all latitudes except close to the poles. We are
currently investigating the magnetic field information close to the
poles and at scales smaller than those provided by IGRF and plan to
report on it in a future publication.

The polarization of the atmosphere should be considered carefully by
ground based experiments that attempt to measure the CMB B mode
polarization signal.  As Fig.~\ref{fig:atm_and_cmb} shows, a 0.01\%
conversion of circular to linear polarization in the instrument gives
rise to an atmospheric polarization signal that is larger by $\sim
\times 2$ than the largest B signal expected with a cosmology that has
a tensor to scalar ratio $r=0.01$. At large angular scales the
atmospheric polarization is larger than estimates for the polarization
signal due to synchrotron \cite{giardino} and Galactic dust
\cite{fink}. Oblique angle reflection of cicularly polarized light
from telescope mirrors can readily generate a 0.01\% conversion to
linear polarization.

Because the column density of oxygen is much smaller at stratospheric
balloon altitudes the signal observed from such a platform is smaller
by about a factor of 1000 making it an unimportant foreground for CMB
polarization experiments.

\begin{figure}[t]
\centerline{\psfig{file=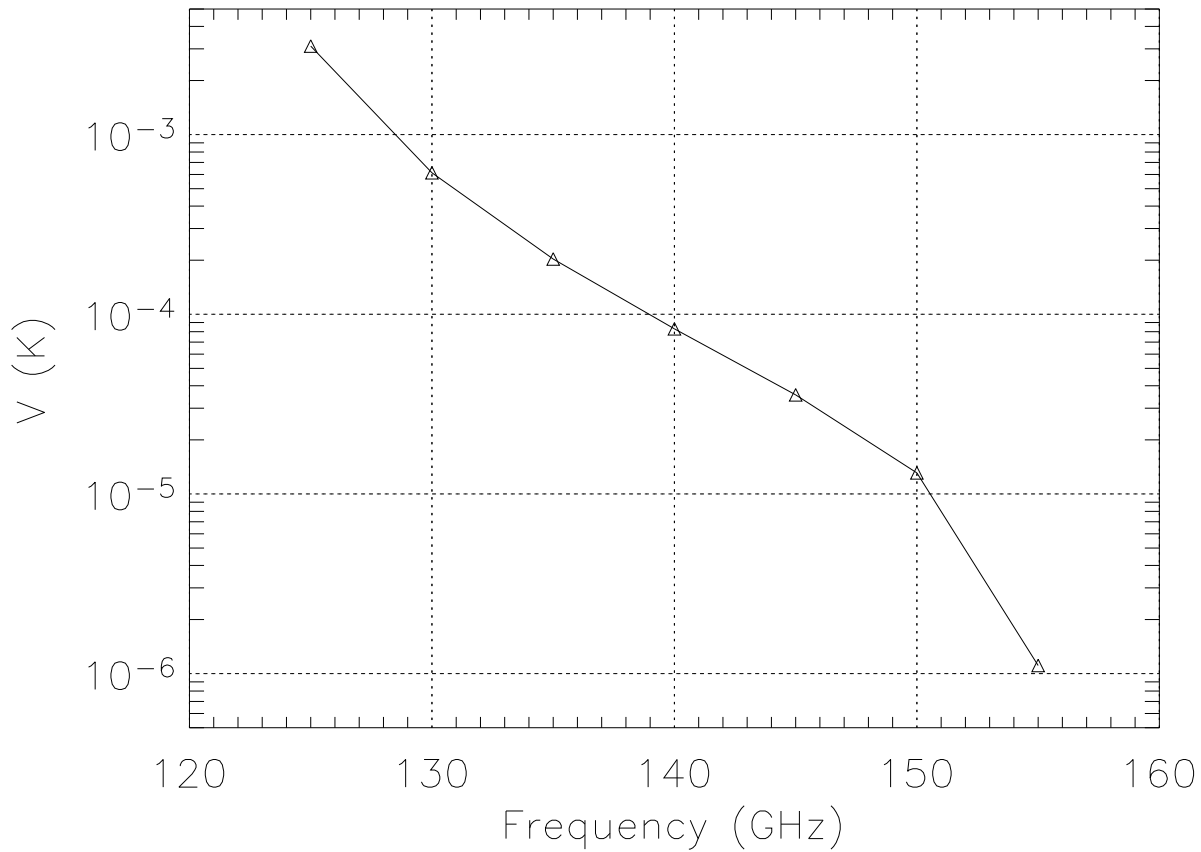,width=5.0in,angle=90}}
\caption{The $V$ Stokes parameter as a function of frequency for 
observation through a US standard atmosphere from an altitude of 1~km.}
\label{fig:freq}
\end{figure}

The atmospheric polarization signals that we have calculated are valid
at a given frequency and have not been integrated over a band of
frequencies; Fig.~\ref{fig:freq} shows the expected variation of the
signal as a function of frequency between 125 and 160~GHz for an
observation at an altitude of 1~km. We do not give detailed
information near the line center at 119 GHz because the line shape is
not accurate enough at these frequencies.  If the magnitude of the
atmospheric polarization at high $\ell$ is the same as that at low
$\ell$ then a strong rejection of out of band power will be necessary
for ground based polarization experiments.

\begin{table}
\caption{Absolute value of the $I$, $Q$, and $V$ 
Stokes parameters for observation through the 
atmosphere in Kelvin (thermodynamic units) for observation sites at two latitudes and two  
altitudes and at the given frequencies. Both sites are at 
a longitude of 140~W and 
the observation is toward the North at an elevation of 45 degrees.}
\begin{center}
\begin{tabular}{cccccc}
\hline
Altitude & Frequency & Latitude & I  & Q & V   \\
  (km)   & (GHz)     & (deg N)  & (K) & (K) & (K)  \\
\hline \hline
 1       & 30        & 40   & 16 & $0.003 \times 10^{-6}$ & $ 76 \times 10^{-6}$ \\
 \cline{3-6}
         &           & 60   & 12 & $0.003 \times 10^{-6}$ & $ 185 \times 10^{-6}$ \\
 \cline{2-6}
         & 90        & 40   & 40 & $0.02 \times 10^{-6}$ & $ 253 \times 10^{-6}$ \\
 \cline{3-6}
         &           & 60   & 25 & $0.02 \times 10^{-6}$ & $ 644 \times 10^{-6}$ \\
 \cline{2-6}
         & 150       & 40   & 81 & $0.003 \times 10^{-6}$ & $ 23 \times 10^{-6}$ \\
 \cline{3-6}
         &            & 60   & 35 & $0.005 \times 10^{-6}$ & $ 78 \times 10^{-6}$ \\
\hline \hline
 30      & 30        & 40   & 2.73 & $0.0005 \times 10^{-9}$ & $ 15 \times 10^{-9}$ \\
 \cline{3-6}
         &           & 60   & 2.73 & $0.0005 \times 10^{-9}$ & $ 28 \times 10^{-9}$ \\
 \cline{2-6}
         & 90        & 40   & 2.73 & $0.005 \times 10^{-9}$ & $ 56 \times 10^{-9}$ \\
 \cline{3-6}
         &           & 60   & 2.73 & $0.005 \times 10^{-9}$ & $ 102 \times 10^{-9}$ \\
 \cline{2-6}
         & 150       & 40   & 2.73 & $0.001 \times 10^{-9}$ & $ 9 \times 10^{-9}$ \\
 \cline{3-6}
         &           & 60   & 2.73 & $0.001 \times 10^{-9}$ & $ 16 \times 10^{-9}$ \\
\hline \hline
\end{tabular} 
\label{tab:std76}
\end{center}
\end{table}

Unlike fluctuations in the brightness temperature of the atmosphere,
which are primarily due to water clouds at low altitudes the polarized
intensity is due to oxygen and is not expected to vary with time,
except that it may be attenuated by varying water vapor. For a given
observation site the signal will be fixed in a particular azimuth and
elevation directions, but not in right ascension and declination.

\section*{Acknowledgements}

We acknowledge use of 
an atmosphere code by J. Pardo and CMBFAST by Seljak and Zaldarriaga. 
We thank M.\ Abroe who helped prepare some of the figures.

\end{document}